\documentstyle[12pt,aaspp]{article}
\newcommand{\etal}{{\it et al.\ }}

\begin{document}

\def\thefootnote{\fnsymbol{footnote}}

\title{THE OPTICAL GRAVITATIONAL LENSING EXPERIMENT. \\
THE CATALOG OF PERIODIC VARIABLE STARS IN THE GALACTIC BULGE. \\
II. PERIODIC VARIABLES IN  FOUR BAADE'S WINDOW FIELDS: BW1, BW2, BW3 and
BW4\footnote{Based on observations obtained at the Las
Campanas Observatory of the Carnegie Institution of Washington}
}

\def\thefootnote{\arabic{footnote}}

\author{A. Udalski\altaffilmark{1,2}, M. Szyma\'nski\altaffilmark{1,2},}
\affil{\tt e-mail I: (udalski,msz)@sirius.astrouw.edu.pl}
\author{J. Ka\l u\.zny\altaffilmark{1,2}, M. Kubiak\altaffilmark{1,2},
M. Mateo\altaffilmark{3}, W. Krzemi\'nski\altaffilmark{4}}
\affil{\tt e-mail I: (jka,mk)@sirius.astrouw.edu.pl,
mateo@astro.lsa.umich.edu, wojtek@roses.ctio.noao.edu}

\altaffiltext{1}{Warsaw University Observatory, Al. Ujazdowskie 4,
00--478 Warszawa, Poland}
\altaffiltext{2}{Visiting Astronomer, Princeton University Observatory,
Princeton, NJ 08544}
\altaffiltext{3}{Department of Astronomy, University of Michigan, 821 Dennison
Bldg., Ann Arbor, MI 48109--1090}
\altaffiltext{4}{Carnegie Observatories, Las Campanas Observatory, Casilla
601, La Serena, Chile}

\begin{abstract}

We present the second part of the OGLE Catalog of Periodic Variable
Stars in the Galactic bulge. 800 variable stars found in four Baade's Window
fields BW1, BW2, BW3 and BW4 are presented. Among them 71 are
classified as pulsating, 465 as eclipsing and 264 as miscellaneous type.
The Catalog and individual observations are available in digital form
from the OGLE Internet archive.

\end{abstract}


\clearpage

\section{INTRODUCTION}

The Optical Gravitational Lensing Experiment (OGLE) is a long term
observing project with the main goal of probing  dark matter in
our Galaxy with microlensing phenomena (Paczy\'nski 1986). After three
seasons of observations toward the Galactic bulge 12 microlensing events
have been detected (Udalski \etal 1993, 1994b, 1994c).

The huge amount of photometric measurements of a few million stars
collected during the OGLE experiment is an unique material for studying
the stellar variability. Large number of detected variable stars, in
some cases significantly exceeding the number of all known variables of
a given type, should considerably contribute to our knowledge  about
origin and evolution of different classes of variable stars and about
structure and evolution of our Galaxy.

The search for periodic variable stars in the OGLE photometric databases
has already been performed and the first part of the OGLE Catalog of
Periodic Variable Stars in the Galactic bulge has been published
(Udalski \etal 1994d; hereafter Paper~I). First part of the Catalog
included 213 periodic variable stars from the center of the Baade's
Window field designated as BWC. This paper is a continuation of the
Catalog --  variable stars from four next fields: BW1, BW2, BW3 and BW4
(Udalski \etal 1994a) are presented in a similar form as  in the first
part of the Catalog.

\section{The CATALOG}

The photometric data presented here were collected during three
observing seasons of the OGLE microlensing search starting from April
13, 1992 through  September 16, 1994. Full logs of observations can be
found in Udalski \etal (1992, 1994a, 1995). Observations were made at
the Las Campanas Observatory, Chile which is operated by the Carnegie
Institution of Washington. The 1-m Swope telescope equipped with
${2048\times 2048}$ Ford/Loral CCD detector was used. Details of data
pipeline,  reduction technique and period search technique can be found
in Paper~I.

Present edition of the Catalog contains periodic variable stars with
${\langle I\rangle}$ brighter than 18~mag.  In the following updates the
Catalog will be extended toward fainter stars. There is also a lower
limit of magnitude: ${I\approx14}$ -- resulting from saturation of
stellar images on CCD frames. The period search was limited to periods
within 0.1 -- 100 days range.

The second part of the Catalog presents variable stars from four Baade's
Window fields: BW1, BW2, BW3 and BW4. Each field covers approximately
${15'\times15'}$ on the sky. Equatorial and galactic coordinates
of these fields are given in Table~1. Each of BW1~--~BW4 fields overlaps
slightly (about ${1'\times1'}$) with the central BWC field. The
variable stars discovered in the overlapping areas during the BWC field
search (Paper~I) are not presented here to avoid ambiguity. In total
17236, 12681, 10779 and 13345 suspected for variability stars were
searched for periodic light variations in fields BW1, BW2, BW3 and BW4,
respectively.

The structure of the Catalog is identical as in the first part
(Paper~I).  Detected variable stars from each field are grouped into
three categories: pulsating stars, eclipsing stars and miscellaneous
type variables. The latter category consists of stars which cannot be
classified unambiguously as pulsating or eclipsing stars. It contains
mostly late type, chromospherically active stars, and likely some
ellipsoidal variables.

For every field and group of stars the Catalog consists of a Table with
basic  parameters for every periodic variable object and an atlas
containing  the  phased light curves and ${30"\times30"}$
finding charts -- part of the  $I$-band frames. North is up and East is
to the left.

The basic parameters for every object include star designation, right
ascention and declination (J2000), period in days and heliocentric
Julian Date of maximum light (minimum for eclipsing variables), $I$
magnitude at maximum brightness, ${V-I}$ color at maximum brightness,
$I$-band amplitude, classification and eventual remarks.

Designation of the object follows the scheme from Paper~I: OGLE {\it
field} V{\it number}, {\it eg.} OGLE~BW2~V22. The variable stars in
every field are initially sorted according to magnitude. Thus lower
number means brighter star.

The equatorial coordinates of variable stars were calculated using
transformation derived from position of stars from the HST Guide Star
Catalog (Lasker \etal 1988). Typically about 20 GSC stars were
identified in each field. Accuracy of coordinates is about $1"$.

Because of strategy adopted in the microlensing search, the vast
majority of measurements was made in the $I$-band (typically 100 -- 190
observations). Only about 10 $V$-band measurements were collected for
each field. Thus, the ${V-I}$ color at maximum light is accurate to 0.05~mag.
In some cases the color is not given -- it could not be derived
because either $V$-band coverage around maximum  was not good enough or
the star could not be identified in the $V$-band database.

Classification within pulsating and eclipsing star groups follows the
scheme of the General Catalog of Variable Stars (Kholopov \etal 1985).

\section{Catalog of Periodic Variable Stars of the BW1, BW2, BW3 and BW4
Fields}

Tables~2~--~13 and Appendices~A~--~L contain the catalog of pulsating,
eclipsing and miscellaneous periodic stars in BW1, BW2, BW3 and
BW4 fields.

71 variable stars were classified as pulsating. Most of them are RR~Lyr
stars type ab and~c. Remaining 18 objects are short period
$\delta$~Scuti type pulsating stars. The periods of majority of them
fall below the lower limit (0.1~day) of period search, but these stars
were identified with ${2\times P}$ period. Thus they probably do not
represent complete sample of this type of stars in the Baade's Window.
Some of RR~Lyr type stars in the Baade's Window were studied by Blanco
(1984). Out of 22 objects from Blanco list located in fields BW1~--~4,
18 were identified in the Catalog as pulsating stars. Two more objects
previously classified as RRc type by Blanco are classified as W~UMa-type
eclipsing variables. Two remaining objects are missing due to reduction
technique drawbacks (see Paper~I). Cross-identification is given in
"Remarks" column of appropriate Tables.

465 eclipsing stars were identified in BW1~--~BW4 fields. The vast
majority (347) of eclipsing objects belongs to W~UMa type (EW). Also 94
Algol-type (EA) and 10 $\beta$~Lyr-type (EB) stars were identified.
14 objects were classified as eclipsing (E) only in ambiguous cases.
Amplitude of some Algol-type eclipsing stars is a lower limit because
the bottom of the eclipse has not been covered. Also the periods of some
Algol variables can be twice of that listed in the Table.

The miscellaneous group of stars contains 264 objects from BW1~--~BW4
fields. Most of them are red giants and subgiants, probably
chromospherically active stars. Some objects in miscellaneous group of
variable stars might be ellipsoidal variables what is indicated in the
"Remarks" column. In such a case the period should be twice of that
given in the Table.

It should be noted that there is a small group of 3 very red (${V-I>5}$)
stars within the miscellaneous group (OGLE~BW3~V1, OGLE~BW3~V2,
OGLE~BW4~V1). These objects have periods longer than 60 days and $I$ amplitude
about 0.3 -- 1~mag. One of these stars -- OGLE~BW4~V1 -- has the period
longer than the upper limit of the period search, and was detected with
shorter period. However it should be  noted that its period is quite uncertain.
The group is highly incomplete because the stars are bright, close to
the saturation level of the  detector, and have long periods.
Nevertheless it distinguishes clearly from other stars of miscellaneous
group.

\newpage

\section{Summary}

We present the second part of the OGLE Catalog of Periodic Variable
Stars in the Galactic Bulge -- periodic variable stars from BW1, BW2,
BW3 and BW4 Baade's Window fields. 800 periodic stars with ${\langle I\rangle}$
brighter than 18~mag. were detected: 71 pulsating, 465 eclipsing and
264 miscellaneous type stars.

The Catalog is supposed to be an open publication and regular updates
are expected when more data become available and search for variables in
fainter objects will be completed. Some errors, unavoidable in this
first release of the Catalog, will also be corrected. Therefore we
expect a feedback from astronomical community when  any errors,
misclassifications etc. are found.

The Catalog and all individual observations of cataloged variable stars
(JD hel., $I$ magnitude, error) are available to astronomical community
from the OGLE Internet archive using anonymous ftp service from
sirius.astrouw.edu.pl host (148.81.8.1), directory
{\it /ogle/var$_-$catalog}.  See README file in this directory.

\acknowledgments{It is a great pleasure to thank B. Paczy\'nski for
valuable suggestions and discussions. This project was supported with
the Polish KBN grant 2P30400306 to A.\ Udalski and  the NSF grants AST
9216494 to B. Paczy\'nski, AST 9216830 to G.W.\ Preston.}

\end{document}